\documentclass{svjour2}                    
\usepackage{graphicx}
\begin{document}

\title{Design and testing of Kinetic Inductance Detectors made of Titanium Nitride
}

\author{P. Diener		\and
		H.G. Leduc		\and
		S.J.C. Yates	\and
		Y.J.Y. Lankwarden	\and	
		J.J.A. Baselmans  
}
\institute{P. Diener 			\and 
			S.J.C. Yates		\and 
			Y.J.Y. Lankwarden	\and 
			J.J.A. Baselmans
\at SRON Netherlands Institute for Space Research, Sorbonnelaan 2, 3584 CA Utrecht, The Netherlands.
\email{p.diener@sron.nl}
\and
H.G. Leduc 
\at Jet Propulsion Laboratory, California Institute of Technology, Pasadena, CA, USA.
}
 \date{   }
\maketitle
\begin{abstract}
To use highly resistive material for Kinetic Inductance Detectors (KID), new designs have to be done, in part due to the impedance match needed between the KID chip and the whole 50 $\Omega$ readout circuit.  
Two new hybrid designs, with an aluminum throughline coupled to titanium nitride microresonators, have been tested experimentally and compared to a similar TiN-only design. With the hybrid designs, parasitic temperature dependent box resonances are absent. The dark KID properties have been measured in a large set of resonators. A surprinsingly long lifetime, up to 5.6 ms is observed in a few KIDs, giving rise to a minimum electrical Noise Equivalent Power of 4.4 $\times$ $ 10^{-20}$ $W\sqrt{Hz}$.

\end{abstract}

\section{Introduction}
\label{sec:intro}
The use of highly resistive material for Kinetic Inductance Detectors is promising because it is expected to increase their responsivity through the high kinetic inductance L$_{s}$ and allow a more efficient direct optical coupling \cite{LeducAPL10}.
To make KIDs based on highly resistive materials, one needs to adapt the microwave design typically used for aluminium KIDs. A KID chip consists essentially of multiple resonators coupled to a microwave throughline, which is connected to the readout electronics. 
The throughline has a characteristic impedance $Z_{0}$ which must match the 50 $\Omega$ impedance of the readout circuit: a mismatch is expected to be at the origin of losses in the transmission and box resonances at the origin of parasitic dips. 
$Z_{0}$ depends in part on the surface inductance L$_{s}$, which is related to the resistivity $\rho$ of the metallic film. A 100 nm thick Al film has typically L$_{s} \sim$ 0.1 pH, whereas L$_{s}\sim$ 44 pH for the TiN film used in this study.
The $Z_{0}$ variations due to the increase of L$_{s}$ can be determined using SONNET electromagnetic 2D simulations. For a typical throughline on silicone substrate having a central line width s = 10 $\mu m$ and a gap width w = 6 $\mu m$, $Z_{0}$ $\approx 53$ $\Omega$ for Al, whereas for L$_{s}$ = 40 pH, $Z_{0}$ becomes frequency dependent, with a mean value at $\approx$ 180 $\Omega$.
 
To solve the problem of impedance mismatch, we have tested the option of an hybrid design, with a throughline in aluminum to read the TiN resonators. Two slightly different designs have been made and measured: the transmission obtained is close to the lossless transmission.The dark KID properties have been studied on a large number of resonators and the best NEP measured is as low as $4.4$ $\times$ $10^{-20}W\sqrt{Hz}$.

\section{Method}
\label{sec:Method}

\begin{figure*}
	\centering
 \includegraphics[width=1\textwidth]{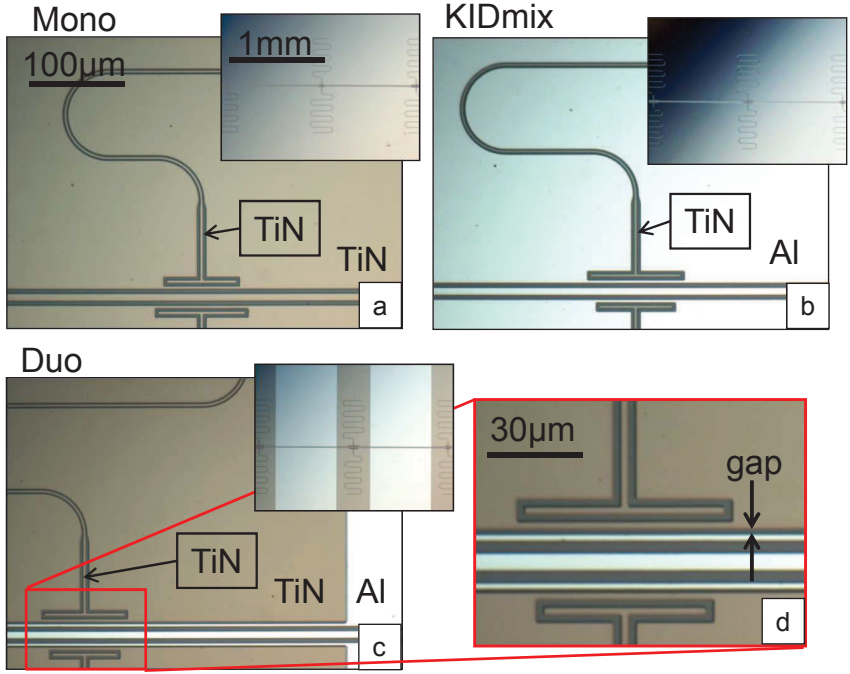}
\caption{a,b,c) photographs of the throughline and one resonator coupler for the three designs studied. The material used for the KID central line and the groundplanes are indicated on each figures. The insets gives a view of each chip in a larger scale. (d) Zoom of the coupling area for the Duo chip: a 2 $\mu m$ gap separates the TiN and Al groundplanes. }
\label{fig:design}       
\end{figure*}

The three designs used for this study are shown in Fig.~\ref{fig:design}.
The first design, Mono, is a standard non hybrid design and has been measured for comparison to see experimentally the effect of an impedance mismatch.
The second design is essentially made of aluminium, with only the KID central lines in TiN. It is called KIDmix because the resonators corresponds to a microwave TiN line having an Al groundplane. With this design, the KID properties may be affected by the presence of aluminium, especially if the superconducting critical temperature T$_{c}$ of TiN is larger than the Al one. Our Al films have T$_{c}$ = 1.2 K and TiN has 0 K $\leq T _{c} \leq$ 4.5 K depending on its stoichiometry. The third design, Duo, has been made for future use of TiN and other high resistive materials having $T_{c} \geq$ 1.2 K. 
It consists of an Al throughline and TiN resonators which are included in rectangular TiN groundplanes.

All designs have a 2 cm length straight throughline, with a central line s = 10 $\mu m$ and a gap width w = 6 $\mu m$. 
The throughline is surrounded by 16 half waves meandered CPW resonators, having a 3 $\mu m$ central line width, a 2 $\mu m$ gap and a length between 3.1 and 4.5 mm. The length sets the resonance frequencies, which we want to be between 4 and 9 GHz to match with our setup specifications. 
It takes into account the geometrical inductance $ L_{g} $ but also the frequency shift $ f_{exp} /f_{g} $ due to the kinetic inductance L$_{s}$.
The resonators are capacitively coupled to the throughline via a T shaped coupler. The coupling length set the coupling quality factor $ Q_{c} $ of each resonator, which is related to the measured (loaded) quality factor Q by the relation $Q^{-1}=Q_{c}^{-1}+Q_{i}^{-1}$ \cite{Mazin_thesis}. 
The internal quality factor $Q_{i}$ is not known a priori so the coupling lengths are between 70 $\mu m$ and 560 $\mu m$ corresponding to $Q_{c}$ between 6 $\times$ $10^{3}$ and 3 $\times$ $10^{6}$. 

In the Duo design, the TiN and Al groundplanes are separated near the throughline by a gap, to avoid the presence of a possibly dirty interface close to the coupler, which might create some excess noise in the KID transmission \cite{note2umgap}. Ideally, the Al groundplane width on both sides of the throughline should be as large as its central line and gaps, to avoid any effect of the TiN groundplanes on $Z_{0}$. But this implies a larger distance between throughline and resonators, thereby reducing $Q_{c}$. To minimize the effect of the TiN groundplanes keeping the desired coupling, the Al groundplane width has been set at 4 $\mu m$ near the resonators and the TiN groundplanes have been reduced to rectangle of 700 $\mu m$ width separated by 1300 $\mu m$ Al groundplanes. 
\\
The 50 nm thick TiN film has been produced by magnetron sputtering onto a highly resistive silicon substrate, with a $N_{2}/Ar$ flow rate of 3.58/30 and patterned using standard contact lithography and dry etching with an SF6/Ar gasmixture.
This is followed by deposition of a 100 nm Al film which is patterned using wet etching.
The TiN film has been characterized by XPS depth profiling and XRD: except on the top 1 nm layer, the film is contaminant free under the sensitivity of 0.5 at$\%$ and the (111) and (200) crystallographic orientations are present, giving rise to diffraction peaks of similar amplitudes.
Resistivity measurements have been made at low temperature on 2 different chips: $ \rho $ $ \approx $ 130 $ \mu\Omega cm $ for both measurement but the superconducting transition has been seen at 0.8 K on the first chip and 1.1 K on the second. 
This discrepancy is unlikely due to the film quality according to the XPS results and may come from the high T$_{c}$ dependence of TiN with nitrogen content, especially around T$_{c}$ = 1 K. 
Indeed, T$_{c}$ = 1.5 K has been measured in another TiN film from the same source and same quality, the only difference between the two films was a change of $ 10^{-3} $ in the ratio of N2/Ar flow rate used during the deposition. 
The chips used in this study were located, on the wafer, near the first chip having $ T_{c} $= 0.8 K which will be used for the following analysis. The superconducting gap $\Delta$ = 125 $\mu eV$ is determined from the relation $\Delta$ = 1.81 k$_{B}$T$_{c}$ \cite{EscoffierPRL04}. The kinetic inductance L$_{s}$ $\approx$ 44 pH is estimated using the relation L$_{s} \approx \hbar \rho/\pi\Delta t$ \cite{LeducAPL10} with t the film thickness.
\\ 
All chips have been measured with the same setup, during separate cooldowns. The chip measured is mounted on a sample holder cooled at 100 mK in an adiabatic demagnetization refrigerator. The sample holder is placed in a light tight box \cite{BaselmansJLTP08}, surrounded by a cryoperm and a superconducting magnetic shield. The complex transmission is measured using a commercial Vector Network Analyzer, the noise is measured using an Agilent synthetizer coupled to an IQ mixer and the lifetime is obtained through the response to an optical pulse. More details on the setup and the measurements can be found ref \cite{BaselmansJLTP08}.

\section{Design comparison}
\label{Results1}

\begin{figure*}
	\centering
	\includegraphics[width=1\textwidth]{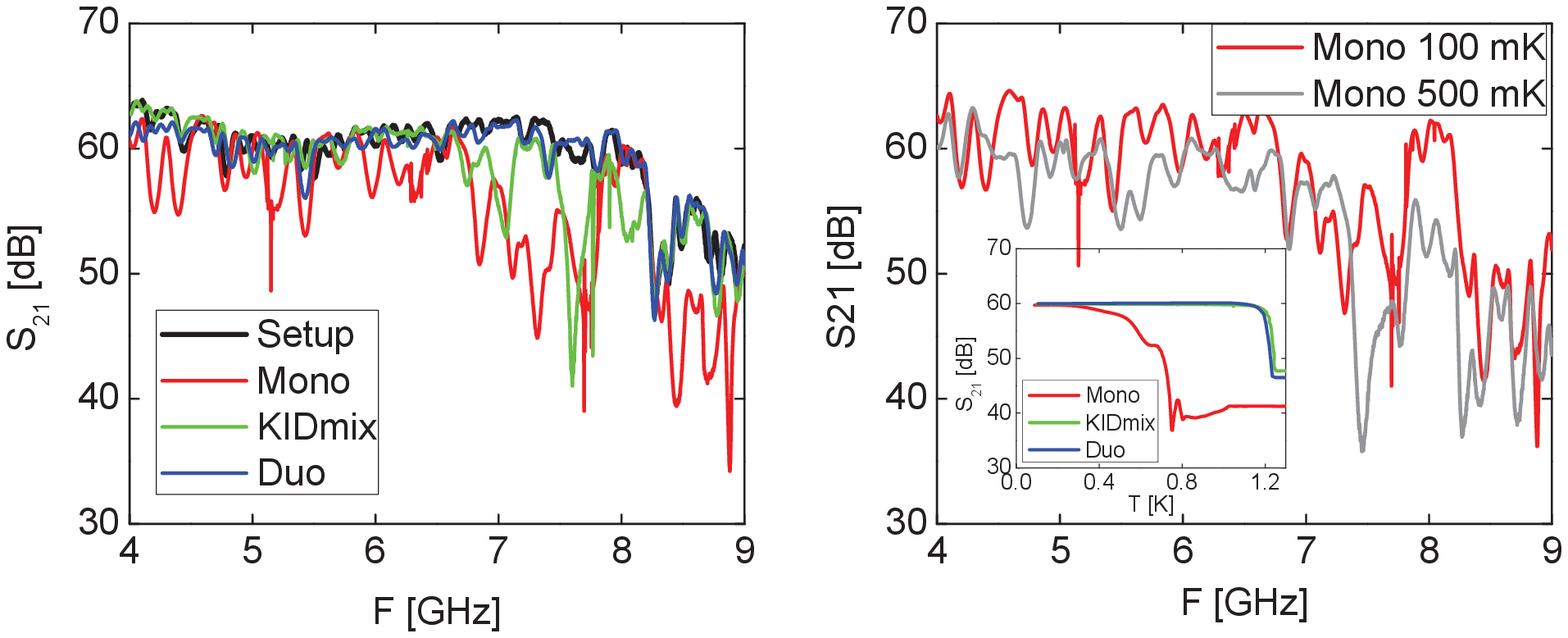}
\caption{(Color online) (a) Magnitude of the complex transmission $S_{21}$ measured at 100 mK on chips having the Mono, Duo and KIDmix designs, compared to an aluminum only chip taken as a reference for the setup noise. (b) $S_{21}$ temperature dependence. F sweep measured at 100 mK and 500 mK for the Mono chip. Inset: temperature sweep at 5 GHz on the Mono and Duo chips. The transmission of Mono exhibits frequency and temperature dependent losses. In contrast, the Duo curve follow closely the lossless transmission. KIDmix has a temperature independent transmission and the shallow dip seen at 7-8 GHz is then attributed to additional setup loss during this measurement.}
\label{fig:S21}       
\end{figure*}

The magnitude of the complex transmission $S_{21}$ measured for the three different designs, at 100 mK and with a readout power of -100 dBm is shown in Fig.~\ref{fig:S21}a. between 4 and 9 GHz. Only few sharp KID dips are visible due to the low frequency band resolution of the sweeps. The result of the same measurement on an aluminum chip is added as a reference for the setup transmission: $S_{21}$ is flat at 60 dB, corresponding to the system transmission and decreases exponentially above 7 GHz due to one amplifier cutoff. \cite{Baselmans_AIP09}. 
The Mono chip exhibits an additional frequency dependent reduction of the transmission, up to 15 dB around 7 GHz. 
On the other hand, the Duo chip curve follow closely the lossless transmission. The KIDmix chip is also close, but shows an unexpected 15 dB reduction in the 7 to 8 GHz region. 

In Fig.~\ref{fig:S21}b., the $S_{21}$ transmission of the Mono chip is presented at 100 mK and 500 mK. The S21 reduction due to the impedance mismatch is frequency dependent but also temperature dependent. For comparison, the inset present the temperature dependence of the transmission, between 100 mK and 1.3K for the three chips at the fixed frequency 5 GHz. For the Duo and KIDmix chips, the curve is dominated by the aluminum transmission: the superconducting transition appear at 1.25 K and $S_{21}$ is almost temperature independent below 800 mK. In the Mono chip, the TiN superconducting transition is seen at 800 mK and the transmission depends on T even below 300 mK. 
The fact that the KIDmix transmission is temperature independent shows that the reduction seen between 7 and 8 Ghz is not due to the presence of TiN. Indeed, it has been identified in following experiments to be a parasitic setup reflection present during the measure of the KIDmix chip: it has been cancelled by retightening one coaxial connector on the 4K stage.

This results show that the throughline impedance match the 50 Ohms circuit impedance for the Duo and KIDmix chips.

\section{KID properties}
\label{Results2}

\begin{table}
	\centering
\caption{Summary of the KID properties obtained for all KIDs: internal quality factor $Q_{i}$, internal power $P_{i}$ in dBm, frequency response $\delta x/\delta N_{qp}$,  phase $S_{\theta}$ and amplitude $S_{R}$ noise at 100 Hz in dBc/Hz, lifetime $\tau$ in ms. The frequency response is obtained from the fit at T $ \geq $ 0.2 K of x=(f(T)-f(0))/f(0) versus $N_{qp}(T)$ calculated with $N_{0}$=8.7 $\times$ $10^{9}eV^{-1}\mu m^{-1}$\cite{LeducAPL10}, for more details see ref \cite{BaselmansJLTP08} .}
\label{tab:1}       
\begin{tabular}{c|c|c|c|c|c|c}
\hline\noalign{\smallskip}
 -- & $Q_{i}$ & $P_{i}$ & $\delta x/\delta N_{qp}$ & $S_{\theta}$  & $S_{R}$  & $\tau$  \\ 
\noalign{\smallskip}\hline\noalign{\smallskip}
Mean value & 1.8 $\times$ $10^{5}$ & -62 & 6.7 $\times$ $10^{-10}$ & -62 & -70 & 1.2 \\
Best value & 4.6 $\times$ $10^{5}$ & -56 & 2.4 $\times$ $10^{-9}$ &  -77 & -83 & 5.6  \\
\noalign{\smallskip}\hline
\end{tabular}
\end{table}

The properties of the microresonators have been measured and analysed following the method described in details in Ref \cite{BaselmansJLTP08}. 
For each chips, multiples dips have been detected ; the measured frequencies of all KIDs in all chips correspond to the fundamental and first harmonic of the geometric frequencies, shifted by a factor fexp/fg=0.2$\pm$0.02. In this study, 5, 13 and 18 KID dips have been fully characterized in, respectively, the Mono, Duo and KIDmix chips.

In the Mono chip, almost all KID dips were asymmetric. This is probably another consequence of the impedance mismatch: the standing waves induce impedance variations near the KIDs coupling arms. It gives rise to an overestimation of the quality factors, but shouldn't affect the determination of the other properties. 
The mean and best values of the KID properties are summarized in Table ~\ref{tab:1} for all KIDs of the three designs. They are varying slightly between KIDs on the same chip and we have not seen any clear systematic properties change between the three designs. 
The mean value for the internal quality factor $Q_{i}$ is 1.8 $\times$ $10^{5}$, which agrees well with the values reported by Vissers et al.\cite{Vissers_APL10} on TiN films having both (111) and (100) XRD peaks: our results then support their conclusions on the dependence of $Q_{i}$ with crystalline orientation. 
The internal powers, calculated for the highest readout power (typically -100dBm) above which an excess noise due to current saturation is observed, are about 10 dB lower than in the case of similar Al KIDs having the same thickness.   

\begin{figure}
	\centering

\includegraphics[width=1\textwidth]{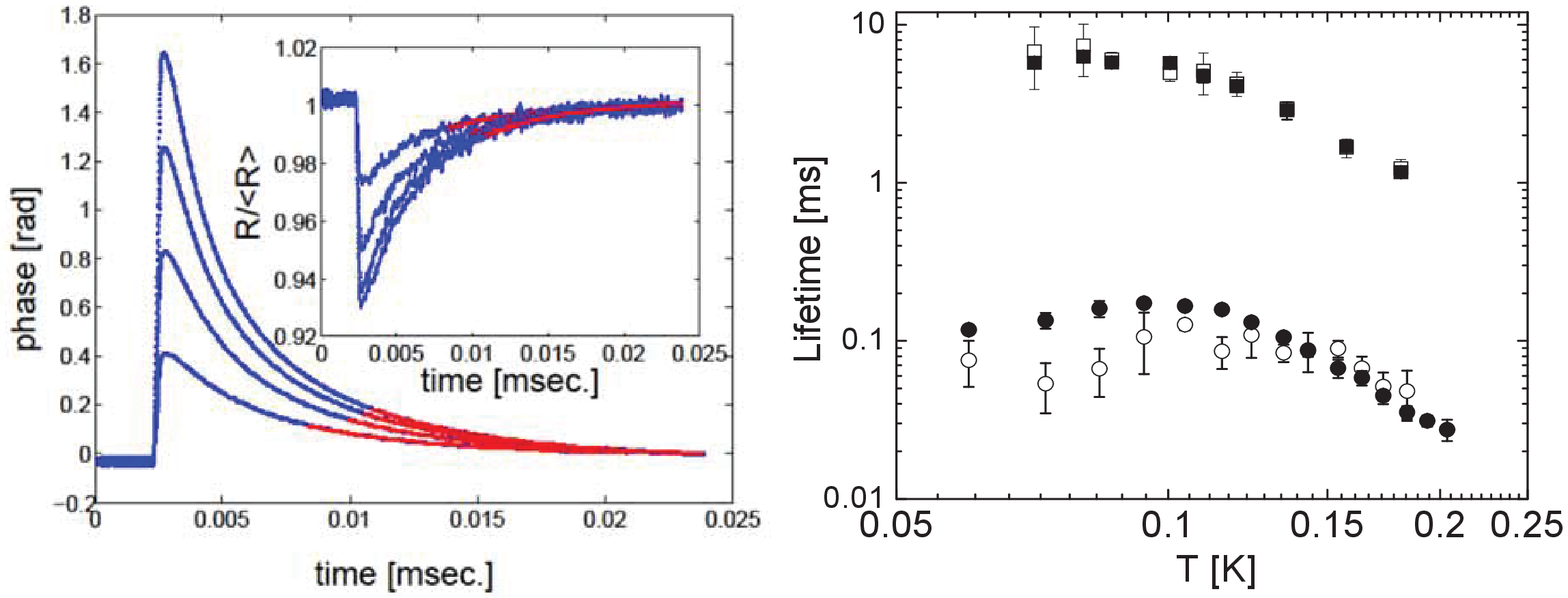}
\caption{Left) (Color online) Phase and amplitude relaxation in KID $\#1$ (DUO chip, f$_{exp}$ $\approx$5.7 GHz, Q$_{i}$ $\approx$2$\times$ $10^{4}$, P$_{i}$ $\approx$-59 dBm) under an optical pulse and for different pulse power (blue). The lifetime deduced from the low angle fits\cite{BaselmansJLTP08} (red) are 5.6 ms and 5.1 ms from resp. the phase and the amplitude. (Right) Temperature dependence of the lifetime in phase (closed symbols) and amplitude (open symbols) for KID $\#1$ (square) and KID $\#2$ (circle) (KIDmix chip, f$_{exp}$ $\approx$3.0 GHz, Q$_{i}$ $\approx$5$\times$ $10^{5}$, P$_{i}$ $\approx$-62 dBm).}
\label{fig:lifetime}       
\end{figure}

On the other hand, the lifetime measured at 100 mK varies widely from one KID to another, even in the same chip. Half of the KIDs (distributed on all the chips) exhibit a lifetime between 0.2 and 0.3 ms, in good agreement with Ref.\cite{LeducAPL10}, but a surprisingly long lifetime has been observed in some KIDs, up to 5.3ms at 100mK. The result of this measurement for the KID $\#1$ having the longer lifetime is presented Fig.~\ref{fig:lifetime}.  
Fig.~\ref{fig:lifetime} also shows the temperature dependence of the lifetime for KID $\#1$ and KID $\#2$ having the typical 0.2 ms lifetime at 100 mK. Both curves have a non monotonic temperature dependence, with a maximum value at T $\approx$ 100 mK $\approx$ 0.13 T$_{c}$. Also, both temperature dependences do not follow the Kaplan theory \cite{KaplanPRB76} above this value.  
However, we believe that the relaxation time measured is the actual recombination time because 1) A careful optimization of the setup has been done to screen/absorb straylight and illuminate the superconducting film only [Jochem LTD14], 2) With the same setup, we have measured on a low noise Al KID, a lifetime from the pulse technique in good agreement with the one determined from the fluctuations in the noise spectra [deVissers PRL10], 3) As shown Fig.~\ref{fig:lifetime} a very similar lifetime value is obtained from both phase and amplitude data, and is independent of the optical pulse power.

The electrical Noise Equivalent Power is calculated following ref \cite{BaselmansJLTP08}. The mean value obtained for the KIDs having the short 0.2 to 0.3 ms lifetime is 5.4 $\times$ $10^{-19}$ $W\sqrt{Hz}$. The NEP being inversely proportional to the lifetime, an even smaller value is obtained in the KIDs having the higher lifetime values and the best NEP is as low as 4.4 $\times$ $10^{-20}$ $W\sqrt{Hz}$.
  
\section{Conclusion}
\label{Conclusion}

Two new microwave hybrid designs have been made in order to read high L$_{s}$ microresonators via an Al throughline.
They have been tested experimentally using a TiN film having L$_{s}\approx$ 44 pH and T$_{c}$ = 0.8 K. 
No impedance mismatch is seen in the $S_{21}$ transmission and the KID dips are symmetrics.
The dark KID properties have been measured in 36 resonators. The lifetime varies from one KID to another, going from 0.2 to 5.7 ms. The long lifetime observed in some KIDs gives rise to a NEP as low as 4.4 $\times$ $10^{-20}$ $W\sqrt{Hz}$.
The origin of the discrepancy in the lifetime values in a high quality TiN film is still open to question, and may be due to the strong dependence of the superconducting properties with stoichiometry in this material.



\begin{thebibliography}{}
\bibitem{note2umgap}
We have measured in the past an excess noise of 20 dB between two analogous hybrid designs having as a main difference an interface close to the couplers.
\end{thebibliography}
\end{document}